\documentclass[doublespacing]{elsart}

\journal{Chemical Physics Letters}

\usepackage{graphics}
\usepackage{amssymb}

\begin{document}

\begin{frontmatter}

% Title, authors and addresses

% use the thanksref command within \title, \author or \address for footnotes;
% use the corauthref command within \author for corresponding author footnotes;
% use the ead command for the email address,
% and the form \ead[url] for the home page:
% \title{Title\thanksref{label1}}
% \thanks[label1]{}
% \author{Name\corauthref{cor1}\thanksref{label2}}
% \ead{email address}
% \ead[url]{home page}
% \thanks[label2]{}
% \corauth[cor1]{}
% \address{Address\thanksref{label3}}
% \thanks[label3]{}

\title{Encoding multiple quantum coherences in non-commuting bases}

\author{C. Ramanathan, H. Cho, P. Cappellaro, G. S. Boutis, D. G. Cory}
% \ead[url]{home page}
\address{Department of Nuclear Engineering, Massachusetts Institute of Technology, 77 Massachusetts Avenue, Cambridge MA 02139}

%\title{}

% use optional labels to link authors explicitly to addresses:
% \author[label1,label2]{}
% \address[label1]{}
% \address[label2]{}

%\author{}

%\address{}

\begin{abstract}
% Text of abstract

Multiple quantum coherences are typically characterised by their coherence number and the number of spins that make up the state, though only the coherence number is normally measured.  We present a simple set of measurements that extend our knowledge of the multiple quantum state by recording the coherences in both the $x$ basis and the usual $z$ basis.  The coherences in the two bases are related by a similarity transformation.   We characterize the growth of the multiple quantum coherences via measurements in the two bases, and show that the rate varies with the coefficient of the driving term in the Hamiltonian.   Such measurements in non-commuting bases provides additional information over the 1D method about the state of the spin system.  In particular the measurement of coherences in a basis other than the usual $z$ basis allows us to study the dynamics of the spin system under Hamiltonians, such as the secular dipolar Hamiltonian, that conserve
$z$ basis coherence number.

\end{abstract}

%\begin{keyword}
% keywords here, in the form: keyword \sep keyword

% PACS codes here, in the form: \PACS code \sep code
%\PACS 
%\end{keyword}
\end{frontmatter}

% main text
\section*{Introduction}

The many-body behaviour of nuclear spins in a solid was described, for a long time, in the language 
of spin thermodynamics \cite{Jeener-1968,Goldman-1970,Wolf-1979}.  This description is, however, essentially static, and the dynamical behaviour of the system  was typically
ignored, or addressed in terms of memory functions \cite{Demco-1975,Mehring-1983}.  With the advent of multiple quantum  NMR techniques 
\cite{Hatanaka-1975a,Pines-1976,Aue-1976,Vega-1976}, multi-spin processes could be described by
multiple spin correlations and multiple quantum (MQ) coherences.  The selective excitation 
and transformation \cite{Warren-1980} of MQ coherences led to a new picture of 
many-body spin dynamics in a dipolar solid \cite{Baum-1985,Munowitz-1987b,Levy-1992,Lacelle-1993}.  The focus in these experiments was on transitions of coherence number, which were observable, rather than the 
number of spins involved in the MQ coherence states.  However, the object of the 
experiment continued to be ``spin counting'' as these MQ experiments were called. These techniques 
have been used to probe the spatial relationships between spins in large 
macromolecules, polymers and crystalline systems, including determining the dimensionality and 
size of localized spin clusters (see \cite{Weitekamp-1983,Munowitz-1987a,Lacelle-1991,Emsley-1994} for reviews).  

In principle, all the spins in a solid are coupled together through their dipolar fields, and the Hilbert space of the spin system is determined by the total number of spins in the sample.   However, at high field and for temperatures above a few degrees Kelvin, it is sufficient to consider a much smaller spin system to predict the NMR spectrum of a solid \cite{Slichter-1990}.  The sample resembles an ensemble of weakly coupled subsystems, within each of which an effective number of coupled spins is postulated to exist.  In equilibrium, at high field, the spin number is one, as $\rho \approx \sum_{i} I_{z}^{i}$.  

In a strong magnetic field ($B_{0}\hat{z}$), an N spin-1/2 system has $2^{N}$ stationary states.  These can be classified according to the magnetic quantum number, $M_{z} = \sum_{i} m_{zi} = (n_{|+1/2>} - n_{|-1/2>})/2 $, where
 $m_{zi} = \pm 1/2$ is the eigenvalue of the $i$th spin in the system, and the energy eigenvalue corresponding to $M_{z}$ is $E_{z} = -\gamma\hbar B_{0}M_{z}$.    For non-degenerate stationary states there are on the order of $2^{2N-1}$ possible transitions between any two levels.  The difference in $M_{z}$ values between the two levels is referred to as the {\em coherence number}.   If the density operator is expanded in the basis of irreducible tensor operators $T_{lm}$,
\begin{equation}
\rho = \sum_{lm} a_{lm} T_{lm}
\end{equation}
the rank $l$ of the tensor element defines the spin number, while the order $m$ characterizes the coherence number.

While these coherences refer to transitions between levels, it is useful to discuss multiple quantum coherences for states of a system. When the state is expressed in the eigenbasis of the system, the presence of a non-zero matrix element $<z_{i}|\rho|z_{j}>$, indicates the presence of an $n$ quantum coherence, where $n = M_{z}(z_{j}) - M_{z}(z_{i})$.   Since $M_{z} = \sum_{i}m_{zi}$ is a good quantum number, we use a collective rotation about the axis of quantization, $\sum_{i} I_{z}^{i}$, to characterize it
\begin{equation}
<z_{i}|\exp(-i\phi\sum_{i}I_{z}^{i})\rho\exp(+i\phi\sum_{i}I_{z}^{i})|z_{j}> = \exp(in\phi) <z_{i}|\rho|z_{j}> \: \: .
\end{equation}

In the usual MQNMR experiment, the system (initially in Zeeman equilibrium) is allowed to evolve under the action of a 
Hamiltonian that generates single quantum (SQ) \cite{Suter-1987}, or double quantum (DQ) \cite{Warren-1980,Yen-1983} 
transitions.  This progressively increases the coherence numbers of the state of the system, as well as causing its spin 
number to increase. 

While the coherences have a physical meaning in the eigenbasis of the system, a generalized coherence number reports on the response of the system to any collective rotation of the spins
(about the $x$ axis for example).  This is equivalent to expressing the state of the spins in a basis where the apparent axis of quantization is given by the axis of rotation, and can be obtained from the eigenbasis via a similarity transformation.  For example, the similarity transform $\mathcal{P}$ connects the density matrices of the system
in the two (the $z$ or eigen-basis, and the $x$ basis) representations.
\begin{equation}
\left[\rho^{x}\right] = \mathcal{P}^{-1} \left[\rho^{z}\right] \mathcal{P}
\end{equation}
where the elements of the matrices are $\left[\rho^{x}\right]_{ij} = <x_{i}|\rho|x_{j}>$, 
$\left[\rho^{z}\right]_{ij} = <z_{i}|\rho|z_{j}>$ and $\{x_{i}\}$ and $\{z_{i}\}$ are 
complete sets of basis operators.  
 Under a collective rotation about the $x$ axis, we obtain, 
\begin{equation}
<x_{i}|\exp(-i\phi\sum_{i}I_{x}^{i})\rho\exp(+i\phi\sum_{i}I_{x}^{i})|x_{j}> = \exp(in\phi) <x_{i}|\rho|x_{j}>
\end{equation}
where $n$ is the $x$ basis coherence number.
Similarity transformations do not change the eigen-energies or the physics of the system \cite{Cornwell-1997}. 
Suter and Pearson \cite{Suter-1988} previously used a combination of phase shifts and a variable flip angle pulse to encode for coherences in the $y$ basis as well as the $z$ basis.   Their technique was recently used to study the dynamics of polarization and coherence echoes \cite{Tomaselli-1996}. Requantization in an alternative basis has also been applied to analyzing RF 
gradient NMR spectroscopy \cite{Zhang-1995}.

In this letter we demonstrate an improved technique for the encoding of coherences in the $x$ basis as well as for encoding coherences simultaneously in the $x$ and $z$ bases.  While the measurement of coherence number in an orthogonal basis does provide more information about the state, it does not yield a direct measure of the spin number.   Under a collective rotation of the spins, the different orders within a given rank are mixed, but there is no mixing between terms of different rank.  Thus contributions to a given coherence
order from the different ranked tensors  cannot be separated out, without some measure of the distribution of tensor ranks in the system.  In order to unambiguosly determine the spin number, N independent measurements are required.

Measurements in non-commuting basis are central to the task of quantum state tomography.  Eigenbasis measurements provide information on the amplitudes of the terms in the density matrix (in the eigenbasis), but not on the associated phase factors.  Changing the basis and repeating the measurements allows reconstruction of the exact state of the system, a familiar process used to measure the Wigner function in optics experiments \cite{Leonhardt-1996}.  Measuring multiple quantum coherences in a basis other than the usual $z$ basis is particularly important if we wish to study the dynamics of the spin system under a Hamiltonian that conserves $z$ basis coherence number, such as the secular dipolar Hamiltonian.

\section*{Methods}

Table 1 shows the initial state, Hamiltonian, and selection rules for the standard MQ 
experiment (using a DQ Hamiltonian), in both the standard $z$ basis and the $x$ basis.
Reference \cite{Zhang-1995} tabulates the transformations between 
quantization in the different Cartesian bases.  Thus, starting from an initial Zeeman state, we see that under the DQ Hamiltonian 
we should get only even order coherences in the $z$ basis and only odd order coherences 
in the $x$ basis. 

\begin{table}
\caption{Description of the MQ experiment in the $z$ and $x$ bases.}

\begin{tabular}{|l|c|c|}  \hline
& $z$ basis & $x$ basis \\ \hline
initial state & $I_{z}$ & $-\frac{\imath}{2}\left\{ I_{x}^{+} - I_{x}^{-}\right\}$ \\ \hline
initial coherence number & $0$ & $\pm 1$ \\ \hline
MQ Hamiltonian & $\displaystyle{\sum_{i<j}} d_{ij} \left\{ I_{i}^{+}I_{j}^{+} +  I_{i}^{-}I_{j}^{-} \right]$ & $\displaystyle{\sum_{i<j}} d_{ij} \left[ \{2I_{xi}I_{xj} - \frac{1}{2}(I_{xi}^{+}I_{xj}^{-}+I_{xi}^{-}I_{xj}^{+})\} - \right.$ \\ 
& & \hspace*{1in} $\left. \frac{1}{2}\{ I_{xi}^{+}I_{xj}^{+} +  I_{xi}^{-}I_{xj}^{-}\} \right\}$ \\ \hline
coherence number selection rule & $\pm2$ & $0,\pm2$ \\ \hline
spin number selection rule & $\pm1$ & $\pm1$ \\ \hline
coherences & even & odd \\ \hline

\end{tabular}
\vspace*{0.5in}

\end{table}

The experimental methods presented here improve on those of Suter and Pearson, as their variable flip angle pulse is replaced by a sequence of phase shifted pulses, whose duration is fixed.  The dipolar evolution during the variable angle pulse, as it is sampled out to multiples of $2\pi$, can significantly attenuate the signal and compromise the resolution of the coherences in the $x$ or $y$ basis, especially in a strongly dipolar coupled system.  In our experiment, the dipolar evolution is refocused, and the $\phi I_{x}$ rotation achieved purely with phase shifts.

The pulse sequences shown in Figures~1(a) and 1(b) allow us to encode coherences in the two bases
under essentially identical conditions.  Figure~1(a) is a $z$ basis encoding experiment while Figure 1(b) is an 
$x$ basis encoding experiment.  The only difference between the two is that the first $\pi/2$ pulse is phase shifted 
along with $U_{\phi}$ in the $x$ basis experiment.  Figure~1(c) shows the 16 pulse DQ
 selective sequence used.  It consists of two cycles of the standard 8 pulse sequence, phase shifted by $\pi$
with respect to each other.  The sequence compensates for pulse imperfections and resonance offsets.  $U_{\phi}$ is created by phase shifting all the pulses in the 16 pulse experiment by $\phi$.
  The two ($\pi/2$) pulses and the Cory 48-pulse sequence are not required for the $z$ basis experiment.  However, they are included in order to perform the two experiments under identical conditions.  In the $x$ basis experiment, the two ($\pi/2$) pulses perform the basis transformation, and the phase encoding of the coherences.  Placing them back-to-back leads to unwanted switching transients, so they are separated by the Cory 48-pulse sequence which prevents evolution of the spin system under the secular dipolar Hamiltonian between the two ($\pi/2$) pulses, and has been described previously \cite{Cory-1990b}.
\begin{figure}
\includegraphics{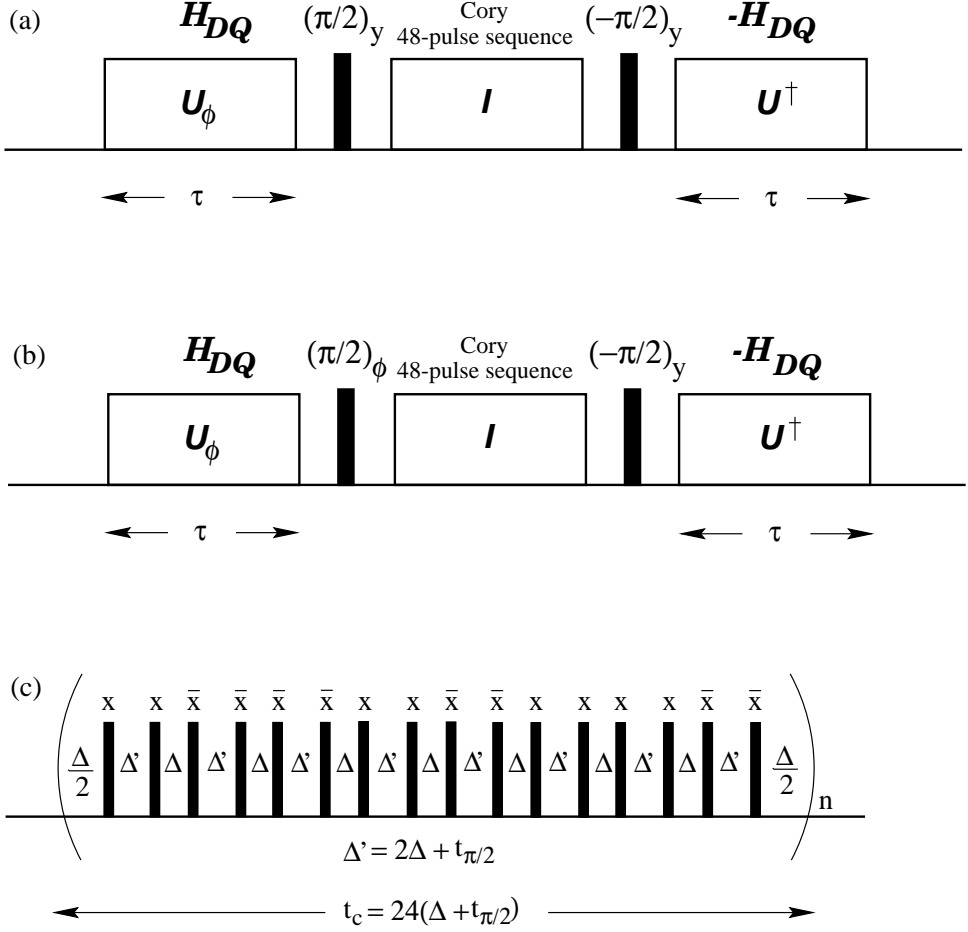}
\caption{ (a) The $z$ basis encoding experiment, $U_{\phi} = R_{z}(-\phi)UR_{z}(\phi)$, and $U = \exp(iH_{DQ}\tau)$.  The propagator for the 48 pulse time-suspension sequence is the Identity Operator I. (b) The $x$ basis encoding experiment where the first ($\pi/2$) pulse is phase shifted, $(\pi/2)_{\phi} = R_{z}(-\phi)(\pi/2)_{y}R_{z}(\phi)$. (c) The 16 pulse sequence used to generate the effective DQ Hamiltonian; $\Delta = 1.3$ $\mu$s, $t_{\pi/2} = 0.51$ $\mu$s, $t_{c} = 43.4$ $\mu$s.}
\end{figure}

The operator corresponding to the observable signal is $I_{z}$.   The measured signal 
for the experiment in Figure~1(a) corresponds to $<I_{z}>_{\phi} = Tr \left[\rho_{f} I_{z} \right]$, 
where the final density matrix is given by  
\begin{eqnarray}
\rho_{f} & = & U^{\dag}R_{y}(-\pi/2)R_{y}(\pi/2)U_{\phi} \rho_{i} U_{\phi}^{\dag}R_{y}(-\pi/2)R_{y}(\pi/2)U \nonumber \\ \nonumber \\
% & = & U^{\dag}U_{\phi} I_{z} U_{\phi}^{\dag}U \nonumber \\ \nonumber \\
% & = & U^{\dag}R_{z}(-\phi)UR_{z}(\phi) I_{z} R_{z}(-\phi)U^{\dag}R_{z}(\phi)U \nonumber  \\ \nonumber \\
 & = & U^{\dag}R_{z}(-\phi)U I_{z} U^{\dag}R_{z}(\phi)U
\end{eqnarray}
where $R_{\alpha}(\phi) = \exp(i\phi I_{\alpha})$, and we have used the fact that the initial state $I_{z}$ is invariant to $z$-rotations.
Defining  $\rho_{s} = U \rho_{i} U^{\dag} = U I_{z} U^{\dag}$, the state of the system after evolution under the
DQ Hamiltonian, we obtain the measured signal in the $z$ basis experiment
\begin{equation}
<I_{z}>_{\phi}  = Tr \left[ R_{z}(-\phi) \rho_{s} R_{z}(\phi) \rho_{s} \right] \label{eq:C} \: \: .
\end{equation}

For the experiment in Figure~1(b), the final density matrix is given by
\begin{eqnarray}
\rho_{f} & = & U^{\dag}R_{y}(-\pi/2)R_{\phi}(\pi/2)U_{\phi} \rho_{i} U_{\phi}^{\dag}R_{\phi}(-\pi/2)R_{y}(\pi/2)U \nonumber \\ \nonumber \\
%& = & U^{\dag}R_{y}(-\pi/2)R_{\phi}(\pi/2)R_{z}(-\phi)UR_{z}(\phi) I_{z} R_{z}(-\phi)U^{\dag}R_{z}(\phi)R_{\phi}(-\pi/2)R_{y}(\pi/2)U \nonumber \\ \nonumber \\
%& = & U^{\dag}R_{y}(-\pi/2)R_{z}(-\phi)R_{y}(\pi/2)R_{z}(\phi)R_{z}(-\phi)UI_{z}U^{\dag}R_{z}(\phi)R_{z}(-\phi)R_{y}(-\pi/2)R_{z}(\phi)R_{y}(\pi/2)U \nonumber \\ \nonumber \\
& = & U^{\dag}R_{y}(-\pi/2)R_{z}(-\phi)R_{y}(\pi/2)U I_{z} U^{\dag}R_{y}(-\pi/2)R_{z}(\phi)R_{y}(\pi/2)U \nonumber \\ \nonumber \\
 & = & U^{\dag}R_{x}(-\phi)UI_{z}U^{\dag}R_{x}(\phi)U
\end{eqnarray}
and the observed magnetization in the $x$ basis experiment is
\begin{equation}
<I_{z}>_{\phi} = Tr \left[ R_{x}(-\phi) \rho_{s} R_{x}(\phi) \rho_{s} \right] \label{eq:D} \: \: .
\end{equation}
In both cases, the experiment is repeated multiple times as $\phi$ is uniformly sampled out to a multiple of $2\pi$, and the resulting data Fourier transformed with respect to $\phi$ to obtain the distribution of coherence numbers.

Note that if Equations~(\ref{eq:C}) and (\ref{eq:D}) could be written in terms of 
$<I_{z}>_{\phi} = Tr \left[ A(\phi) \rho_{s} \right] $,
where the set of operators $A(\phi)$ form a complete basis for the Hilbert space of the spin 
system, it would be possible to perform quantum state tomography on the spins 
\cite{Leonhardt-1996}.  However, the experiments described here cannot completely characterize the state, or even
just its collective properties.

\section*{Results}

The experiments were performed at 2.35 T with a Bruker Avance 
spectrometer and a home-built RF probe.  The 90 degree pulse time was 0.51 $\mu$s.  The pulse spacing $\Delta$ used in 
the DQ sequence was 1.3 $\mu$s, and the cycle time for the 16 pulse cycle was 43.4 $\mu$s.
The pulse spacing in the 48 pulse time suspension sequence was 1.5 $\mu$s.
The T$_{1}$ of the single crystal calcium fluoride sample used was 7 s, and the recycle delay used in the experiment was 10 s. 

Figure~2 shows the results obtained in the $z$ and $x$ basis encoding experiments.   The maximum coherence encoded was $\pm32$, with $\Delta\phi = 2\pi/64$.  The phase incrementation 
was carried out to $8 \pi$.  
It is seen that the $z$ and $x$ basis measurements give only even and odd coherences
respectively as expected from Table~1. The data shown correspond to 1, 3 and
5 loops of the 16 pulse cycle. The higher order coherences are seen to grow in both bases, as the system evolves under the DQ Hamiltonian. 

\begin{figure}
\includegraphics*{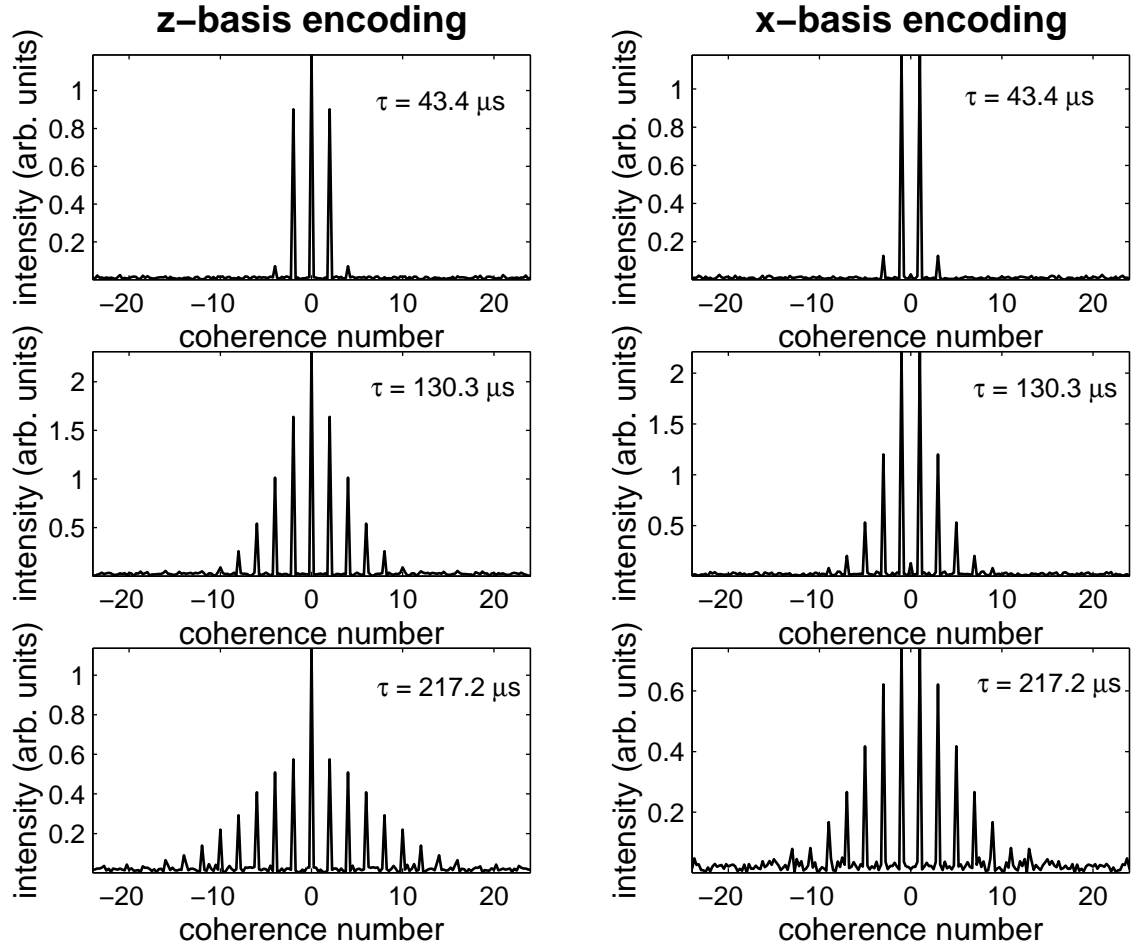}
\caption{ Comparison of $z$ basis and $x$ basis coherences at preparations times $\tau =$ 43.4, 130.3 and 217.2 $\mu$s, corresponding to 1, 3 and 5 loops of the 16 pulse cycle in Figure 1(c), showing the presence of purely even and odd coherences in the two bases respectively.}
\end{figure}

In Figure~3 we plot the effective spin number, obtained by a Gaussian fit to the coherence number distributions in the $x$ and $z$ bases ($N(\tau) = 2\sigma^{2}$) as a function of $\tau$ \cite{Baum-1985}.   The fits were performed on the 1D data.  The variance of the $x$ basis measurements is consistently smaller than that of the $z$ basis measurements. 
The points appear to lie on a straight line, and a best linear fit has a slope of 0.54, which is very close to the value of 0.5 expected from the ratio between the double quantum selective terms in the DQ Hamiltonian expressed in the two bases, as shown in Table 1.  The linear fit was performed on the mean spin numbers $N_x$ and $N_z$ without considering the standard deviations. The error in the fit is negligible.  In the $z$ basis, evolution under the DQ Hamiltonian forces the system to change coherence number, while in the $x$ basis the presence of zero quantum terms permits mixing without changing the coherence number.  Thus the growth of the coherence numbers is slowed relative to the $z$ basis, leading to a narrower distribution.   It should be emphasised that the basis change does not change the spin number, only the experimentally observable coherences.  The change in spin number with basis representation demonstrates the limitations of the Gaussian statistical model as an accurate predictor of spin number in strongly coupled spin systems.
Lacelle has also discussed the limitations of the model \cite{Lacelle-1991}.

\begin{figure}
\includegraphics*{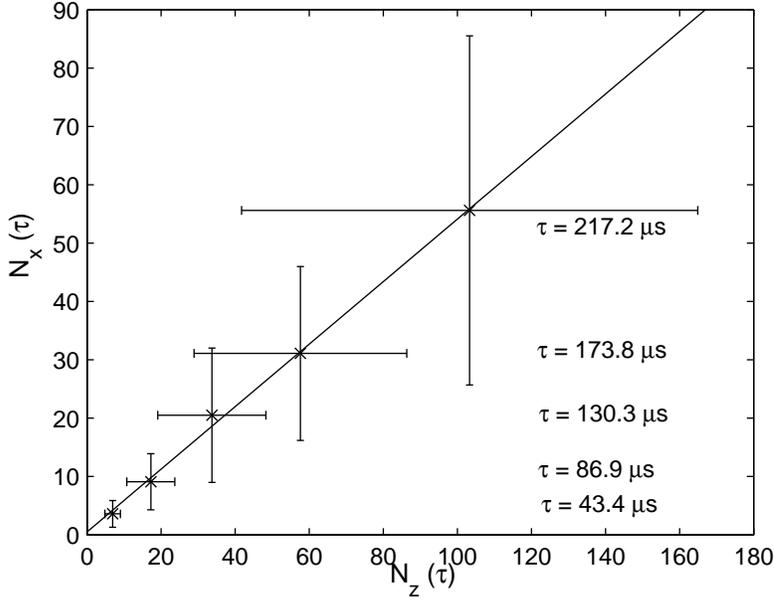}
\caption{ Plot of the effective spin number ($N(\tau) = 2\sigma^{2}$) obtained by fitting the coherence distributions obtained in the 1D $x$ and $z$ basis measurements to a Gaussian.  Also shown is the best linear fit to the data, whose slope is 0.54.  The slope expected from the coefficient of the DQ selective terms in the two bases (see Table 1) is 0.5.}
\end{figure}

As first shown by Suter and Pearson \cite{Suter-1988}, a two dimensional experiment illustrates the 
correlation between $x$ and $z$ basis coherences.   The two-dimensional experiment is obtained from the $x$ basis experiment in Figure 1(b) by phase cycling the refocussing sequence $U^{\dag}$ by $\beta$ independently of $\phi$.
The phases $\phi$ and $\beta$ are incremented independently to sample a rectangular grid and a 2D Fourier transform is performed to yield the coherences. The measured data in a single experiment is
\begin{equation}
<I_{z}>_{\beta\phi} = Tr \left[R_{z}(-\beta)R_{x}(-\phi) \rho_{s} R_{x}(\phi)R_{z}(\beta) \rho_{s} \right] \label{eq:E} \: .
\end{equation}
It is straightforward to show that the order of the $x$ and 
$z$ phase shifts does not matter when both of them are sampled over a $2\pi$ range.  
The $(\phi_{x})(\beta_{z})$ experiment is equivalent to the $(-\beta_{z})(-\phi_{x})$ 
experiment.  The two dimensional experiment separates out the different terms that contribute to a particular $z$ basis coherence as can be seen in Figure 4.
We used $\tau$ = 130.3 $\mu$s, corresponding to 3 loops of the MQ cycle. The maximum coherence encoded in each direction was $\pm12$, with $\Delta\phi = 2\pi/24$.  The phase incrementation was carried out to $8 \pi$ along each axis, resulting in a $96\times96$ data grid, which was Fourier transformed to yield the coherences shown.

\begin{figure}
\includegraphics*{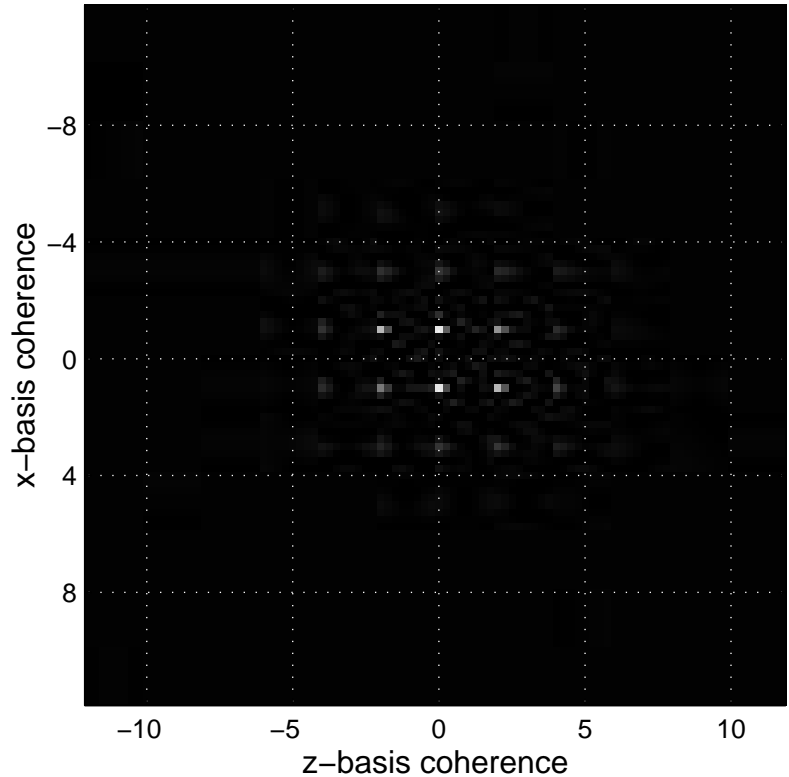}
\caption{Results of the 2D experiment - showing correlations between encoding in the $x$ and $z$ bases.  The preparation time used was $\tau$ = 130.3 $\mu$s, corresponding to 3 loops of the 16 pulse cycle in Figure 1(c). The width in $z$ appears broader than the width in $x$.}  
\end{figure}

This two dimensional technique can be used to examine the evolution of multiple quantum coherences under the dipolar Hamiltonian.  Figure 5 shows the attenuation of the $z$ basis zero quantum signal as it evolves under the dipolar Hamiltonian.  Also shown on the figure are the various $x$ basis contributions to the $z$ basis zero quantum signal obtained from the 2D data.  The decay is clearly non-exponential.  It can be seen that the different $x$ basis terms attenuate at different rates, and that the measured decay of the 1D $z$ basis data represents some sort of average attenuation of all these terms.  Thus this technique allows us to probe the details of spin dynamics beyond the ability of existing techniques.  We have also recently used this technique to study the evolution of the spin system following a Jeener-Broekaert pulse pair, and observed the evolution of the system to a dipolar ordered state \cite{Cho-2002}.

\begin{figure}
\includegraphics*{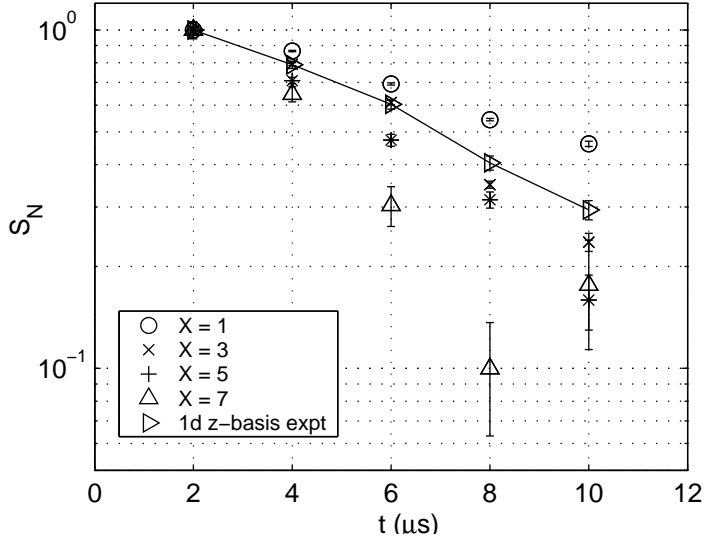}
\caption{Decay of the $z$ basis zero quantum signal under the dipolar Hamiltonian.  The decay of the different $x$ basis contributions to the zero quantum signal in the $z$ basis, obtained from the 2D experiment are also shown.  The data were normalized to the observed intensity at 2 $\mu$s.  Note that the data are plotted on a log scale.  The preparation time used was $\tau$ = 130.3 $\mu$s, corresponding to 3 loops of the 16 pulse cycle in Figure 1(c).}  
\end{figure}

\section*{Conclusions}
We have shown that by encoding MQ coherences in different bases ($x$ and $z$) additional information about the state of the spin system may be obtained.  In particular, $x$ basis encoding could be useful in determining the size of multiple spin correlations under the action of a Hamiltonian that preserves $z$ basis coherence number, but changes the number of spins in the state, such as the secular dipolar Hamiltonian.

\section*{Acknowledgements}
We would like to thank Dr.\ Joseph Emerson for stimulating discussions, and the NSF and DARPA DSO for financial support.

\bibliography{Bibliography}

\newpage

\section*{Figure Captions}
\begin{enumerate}

\item (a) The $z$ basis encoding experiment, $U_{\phi} = R_{z}(-\phi)UR_{z}(\phi)$, and $U = \exp(iH_{DQ}\tau)$.  The propagator for the 48 pulse time-suspension sequence is the Identity Operator I. (b) The $x$ basis encoding experiment. where the first ($\pi/2$) pulse is phase shifted, $(\pi/2)_{\phi} = R_{z}(-\phi)(\pi/2)_{y}R_{z}(\phi)$. (c) The 16 pulse sequence used to generate the effective DQ Hamiltonian;  $\Delta = 1.3$ $\mu$s, $t_{\pi/2} = 0.51$ $\mu$s, $t_{c} =  43.4$ $\mu$s.

\item Comparison of $z$ basis and $x$ basis coherences at preparations times $\tau =$ 43.4, 130.3 and 217.2 $\mu$s, corresponding to 1, 3 and 5 loops of the 16 pulse cycle in Figure 1(c), showing the presence of purely even and odd coherences in the two bases respectively.

\item Plot of the effective spin number ($N(\tau) = 2\sigma^{2}$) obtained by fitting the coherence distributions obtained in the 1D $x$ and $z$ basis measurements to a Gaussian.  Also shown is the best linear fit to the data, whose slope is 0.54.  The slope expected from the coefficient of the DQ selective terms in the two bases (see Table 1) is 0.5.

\item Results of the 2D experiment - showing correlations between encoding in the $x$ and $z$ bases.  The preparation time used was $\tau$ = 130.3 $\mu$s, corresponding to 3 loops of the 16 pulse cycle in Figure 1(c).

\item Decay of the $z$ basis zero quantum signal under the dipolar Hamiltonian.  The decay of the different $x$ basis contributions to the zero quantum signal in the $z$ basis, obtained from the 2D experiment are also shown.  The data were normalized to the observed intensity at 2 $\mu$s.  Note that the data are plotted on a log scale.  The preparation time used was $\tau$ = 130.3 $\mu$s, corresponding to 3 loops of the 16 pulse cycle in Figure 1(c).

\end{enumerate}

\end{document}